\documentclass[12pt,preprint]{aastex}

\def\ds{\displaystyle}

\begin{document}

\title{SELFGRAVITATING GAS SPHERES IN A BOX AND RELATIVISTIC CLUSTERS:
RELATION BETWEEN DYNAMICAL AND THERMODYNAMICAL STABILITY}

\author{Gennady S. BISNOVATYI-KOGAN\altaffilmark{1,2}}
\altaffiltext{1}{Space Research Institute (IKI)
\\ Profsoyuznaya 84/32, Moscow 117997, Russia
\\E-mail: gkogan@iki.rssi.ru}
\altaffiltext{2}{Joint Institute of Nuclear Researches, Dubna, Russia}

\and

\author{Marco MERAFINA\altaffilmark{3}}
\altaffiltext{3}{Department of Physics, University of Rome ``La Sapienza"
\\Piazzale Aldo Moro 2, I-00185 Rome, Italy
\\E-mail: marco.merafina@roma1.infn.it}

\begin{abstract}
We derive a variational principle for the dynamical stability
of a cluster as a gas sphere in a box. Newtonian clusters are always
dynamically stable and, for relativistic clusters, the relation between
dynamical and thermodynamical instabilities is analyzed. The boundaries
between dynamically and thermodynamically stable and unstable models are
found numerically for relativistic stellar systems with different cut off
parameters. A criterion based on binding energy curve is used for
determination of the boundary of dynamical stability.
\end{abstract}

\keywords{dense matter --- galaxies: star clusters --- hydrodynamics ---
instabilities --- relativity}

\section{Introduction}
It is well known that Newtonian stellar clusters with effective adiabatic
power $\gamma =5/3$ are always dynamically stable (see Zel'dovich \&
Novikov 1971). An isolated cluster, in which temperature tends
to a constant value all over the radius, suffers gravothermal catastrophe
(see Antonov 1962; Lynden-Bell \& Wood 1968) at which
any finite object evolves into a model with highly concentrated core and
very extended envelope, where radius and central density tend to infinity.
If we remove a demand of constant temperature and consider a rapid
dynamical (adiabatic) perturbation, then the cluster does not react
drastically and returns always to its mechanical equilibrium.

An interesting analysis of dynamical and thermodynamical instability of
stellar clusters made by Chavanis (2002a, 2002b) contains some sections
that show this problem is still not quite clear, instead of several
publications devoted to this topic (see Merafina 1999; Lightman \& Shapiro
1978).

In section 2 we derive in Newtonian gravity the variational principle for
investigation of dynamical stability of a stellar cluster or a gas in a
spherical box and discuss properties of trial functions which may be
used for a stability analysis. For Newtonian gravity, we find conditions
$\gamma >4/3$ for stability of an extended cluster with $\rho_e/\rho_0\ll
1$ and the condition $\gamma >0$ for stability of a very hot body in a
box at $P/\rho \gg GM/R$ with almost constant pressure $P$ and density
$\rho$. The pressure along the adiabate is supposed to follow the relation
$P=k\rho^{\gamma}$.

Relations between dynamical and thermodynamical stability of relativistic
clusters are analyzed in sections 3 and 4. Numerical results about stability
analysis of relativistic stellar clusters with different cut off parameters
are represented in section 5.

The oscillatory behavior of the mass $M$ of the cluster as a function of a
central density $\rho_0$, at fixed temperature $T$, indicates increasing
number of thermodynamically unstable modes. Similar oscillatory
dependence of $M(\rho_0)$, at fixed parameter $W_0$, shows (approximately)
increasing number of dynamically unstable modes. Dynamically stable models
with arbitrarily large central redshifts $z_c$ exist only at temperature
$T\lesssim 0.06$. All such models are thermodynamically unstable.

\section{Newtonian clusters and gas spheres}
Let us consider development of dynamical perturbations in the cluster,
where characteristic time is usually much shorter than any other time,
including the time of energy exchange with an outer thermostat. In
dynamical (rapid) perturbations, where local entropy is conserved and
there is no time to smooth the temperature, the relations between $\delta
T$, $\delta\rho$ and $\delta P$ follow adiabatic relations

\begin{equation}
{\delta P\over P}={5\over 3}{\delta\rho\over\rho}={5\over2}{\delta
T\over T}~,
\label{eq:1}
\end{equation}
\\
so the pressure perturbation, where in mechanical equilibrium we have

\begin{equation}
{dP\over dr}= {kT\over m_s}{d\rho\over dr}~,
\label{eq:2}
\end{equation}
\\
should be taken as

\begin{equation}
\delta P = \gamma{P\over\rho}\delta\rho~~~~~({\rm where}~\gamma =5/3)~.
\label{eq:3}
\end{equation}
\\
The dynamical stability analysis (Chandrasekhar 1964; Bisnovatyi-Kogan 2001)
is usually performed for a star with zero density and pressure on the
boundary. For a cluster in a box with nonzero $\rho$ and $P$ at the edge,
the boundary conditions are different, but the expression for frequency,
following from the variational principle, remains the same. It is easy to
show, using equations in Lagrangian coordinates in which dynamical
equations at spherical symmetry are written as

\begin{equation}
{\partial v\over \partial t}+{1\over\rho}{\partial P\over\partial r}+{Gm\over
r^2}=0
\label{eq:4}
\end{equation}

\begin{equation}
{\partial r\over \partial t}=v
\label{eq:5}
\end{equation}

\begin{equation}
{\partial r\over \partial m}= {1\over 4\pi\rho r^2}~,
\label{eq:6}
\end{equation}
\\
that, for linear perturbations of the static model with $\delta r\sim
e^{i\sigma t}$, we obtain, using Eq.~(\ref{eq:3}),

\begin{equation}
-\sigma^2\delta r + \delta\left[{{1\over\rho}{d P\over d
r}}\right]-{2Gm\over r^3}\delta r =0
\label{eq:7}
\end{equation}

\begin{equation}
{d\delta r\over dm}= -{1\over 2\pi\rho r^2}{\delta r\over r}-{1\over
4\pi\rho r^2}{\delta\rho\over\rho}~, ~~~~~{\delta\rho\over\rho}=
-2{\delta r\over r}-{d\delta r\over dr}
\label{eq:8}
\end{equation}

\begin{equation}
\delta\left[{1\over\rho}{d P\over d r}\right]=
\delta\left[4\pi r^2{d P\over d m}\right]=
{2\over\rho}{dP\over dr}{\delta r\over r}+{\gamma\over\rho}{d\over dr}
\left({P{\delta\rho\over\rho}}\right).
\label{eq:9}
\end{equation}
\\
The equation for a perturbation $\delta r(m)$, taking into account
Eqs.~(\ref{eq:7}),~(\ref{eq:8}) and the equilibrium equations, is
written as

\begin{equation}
-\sigma^2\delta r+{4\over\rho}{dP\over dr}{\delta r\over r}
-{\gamma\over\rho}{d\over dr}\left[{P\left({2{\delta r\over r}+{d\delta r
\over dr}}\right)}\right] =0~.
\label{eq:10}
\end{equation}
\\
The Eq.~(\ref{eq:10}) for a cluster in a box should be solved at boundary
conditions

\begin{equation}
\delta r(0)=\delta r(M)=0~,~~~~~r(M)=R~,
\label{eq:11}
\end{equation}
\\
with a variable outer density and pressure $\delta\rho (M)\not= 0$,
$\delta P(M)\not= 0$. This is different from the boundary conditions of
stellar oscillations which, in adiabatic approximation, are

\begin{equation}
\left\{
\begin{array}{ll}
\delta r(0) = 0 & \\ & \\
\delta\rho (M)= {\ds -\rho\left({2{\delta r\over r}+{d\delta r\over
dr}}\right)\Biggl\vert_{m=M} =0~,} &
\end{array}\right.
\label{eq:12}
\end{equation}
\\
with a variable outer boundary radius $\delta r(M)\not= 0$.

Instead of solving Eq.~(\ref{eq:10}), let us derive a variational
principle, which gives a possibility of a simple stability
analysis. Integrating Eq.~(\ref{eq:10}) over the mass of the cluster,
after multiplying by $\delta r$, and accepting the normalization of the
linear perturbation function in the form

\begin{equation}
\int _0^M {\delta r^2 {\rm d}m} =A~,
\label{eq:13}
\end{equation}
\\
we obtain from Eq.~(\ref{eq:10}), after partial integration with boundary
conditions (\ref{eq:11}), the following expression for the squared
frequency

\begin{equation}
\sigma ^2 = {1\over A} \int _0^M {{P\over\rho}\left[{\gamma\left({2{\delta
r\over r}+{d\delta r\over dr}}\right)^2 -4\left({\delta r\over r}\right)^2
-8{\delta r\over r}{d\delta r\over dr}}\right] {\rm d}m}~.
\label{eq:14}
\end{equation}
\\
This is exactly the same expression which takes place for a star with
another boundary conditions (\ref{eq:12}).

Variational principle over-estimates the values of $\sigma ^2$ for
different trial functions $\delta r(m)$, giving the minimal exact value
for the eigenfunction of the oscillations. So, variational principle may
prove the existence of instability, but only approximately permits to make
a judgement about the stability of the system. It is important to use only
those trial functions which satisfy the boundary conditions of the
eigefunction.

It is known for stars with boundary conditions (\ref{eq:12}), that the
linear trial function gives almost an exact result for stability boundary.
We have from Eq.~(\ref{eq:14}) for $\delta r =\alpha r$

\begin{equation}
\sigma ^2 = {9\alpha^2\over A} \int _0^M {{P\over\rho}\left({\gamma - 4/3}
\right) {\rm d}m}
\label{eq:15}
\end{equation}
\\
that corresponds to stability boundary at $\gamma =4/3$. This boundary is
the exact value for stars with constant $\gamma$ because at $\gamma =4/3$
the trial function $\delta r =\alpha r$ is also an exact eigenfunction
(Zel'dovich \& Novikov 1971).

For a cluster in the box the linear eigenfunction is not valid, because it
does not satisfy outer boundary conditions. Nevertheless, for
clusters with low ratio $\rho_e/\rho_0$ (where $\rho_e$ and $\rho_0$ are
the external and central density, respectively) the exact fulfilment of
the outer boundary condition is unimportant and, at $\rho_e/\rho_0\to 0$,
the condition (\ref{eq:15}) is approximately valid also as a criterion for
dynamical stability of a cluster in a box. On the other hand, the dynamic
response of a stellar cluster with quasi-maxwellian distribution function
to global radial perturbations is similar to the response for adiabatic
perturbations in a star with the same adiabatic index, which is
$\gamma =5/3$ in the nonrelativistic cluster. In fact, for nonrelativistic
spherical stellar clusters, we obtain dynamical stability because also
barotropic stars with the same density distribution function are dynamically
stable (Antonov 1960). This result leads in many cases to choose a parallel
treatment in stability analysis for these physically different systems
(Binney \& Tremaine 1987). Similar correspondence exists even in the
investigation of thermodynamical stability (Bettwieser \& Sugimoto 1985).
Therefore, nonrelativistic clusters with $\gamma =5/3$ in the process of
gravithermal catastrophe are, always, dynamically stable. Chavanis (2002a)
considered perturbations at constant temperature and obtained dynamic
instability in this case. We should stress that perturbations developing
in the dynamical time do not preserve constant temperature, which could be
reached only by contact with an external thermostat. Therefore the
increment of ``dynamic" instability at constant temperature is determined
by the time of heat exchange between the cluster and thermostat, but not
by the time of dynamical processes inside the clusters.

For clusters in a smaller box the outer boundary condition plays more
important role and, for small boxes, where the gravity is less important
(in the limit $P={\rm constant}$), using the trial function $\delta
r=\alpha r$ gives an highly inadequate result. Using a trial function with
a correct boundary condition is necessary here, what, as evidently
expected, reduce the demand on $\gamma$ for stability of the object in the
box. Therefore, only a relativistic gas in the box may become dynamically
unstable, because the Newtonian clusters have always $\gamma =5/3$. Using
a trial function for a box with radius $R$ and constant pressure $P_0$ as

\begin{equation}
\delta r=\alpha r(1-r/R)~,
\label{eq:16}
\end{equation}
\\
we get from Eq.~(\ref{eq:14})

\begin{eqnarray}
\sigma^2 & = & {4\pi\alpha^2\over A} P_0 \int_0^R {\left[{\gamma~(3-4r/R)^2
-4(1-r/R)^2-8(1-r/R)(1-2r/R)} \right] r^2 {\rm d}r}=
\nonumber \\ & = & {4\pi\alpha^2\over A} P_0 R^3 \left[{\gamma\left({9\over 3}
-{24\over 4}+{16\over 5}\right) -{12\over 3}+{32\over 4}-{20\over 5}}\right] =
{4\pi\alpha^2\over 5A} \gamma P_0 R^3~,
\label{eq:17}
\end{eqnarray}
\vspace{3pt}
\\
where $\gamma$ is constant all over the configuration.

The stability criterion $\gamma >0$ corresponds to the dynamical stability
of a weakly gravitating gas in the box under the condition of a high
pressure $P_0/\rho\gg GM/R$. The criterion $\gamma >0$ remains valid for a
more general trial function $\delta r=\alpha r^n (1-r/R)$, for any value
of $n>0$. We see that all instabilities of the Newtonian isothermal
clusters with different boundary conditions considered by Antonov (1962),
Lynden-Bell \& Wood (1968) and Chavanis (2002a) are related only to
thermodynamical instabilities, being the clusters always dynamically
stable.

\section{Relativistic models of selfgravitating systems}
\centerline{\it\noindent Gaseous systems in a box}

The relation between dynamical and thermodynamical stability is much
tenser in the relativistic case, when increased gravitational force may
lead to dynamical instability for any equation of state.

The analysis made by Chavanis (2002b) demonstrates the
development of dynamical instability for stars with ultrarelativistic
equation of state in the box at increasing box size. We should clearly
distinguish between two limits of the ultrarelativistic equation of state

\begin{equation}
P=q\epsilon
\label{eq:18}
\end{equation}
\\
considered by Chavanis.

The first one corresponds to the deep interior of the neutron star where,
in conditions of ultrarelativistic degeneracy and strong nuclear forces,
we have Eq.~(\ref{eq:18}) with $1/13 < q <1$ (see Ambartsumyan \& Saakyan
1961; Zel'dovich 1962a). The parameter characterizing
the dynamical stability is the adiabatic index (see Harrison et al. 1965)

\begin{equation}
\gamma = {P+\rho c^2\over Pc^2}\left({\partial P\over\partial\rho}\right)_S
= q+1~,
\label{eq:19}
\end{equation}
\\
where, for dynamically stable configurations, we have

\begin{equation}
\gamma > \gamma_{cr}~.
\label{eq:20}
\end{equation}
\\
In Newtonian limit we have always $\gamma_{cr}=4/3$ but, when the effects
of general relativity make the gravity stronger, $\gamma_{cr}$ becomes
larger than $4/3$ (see Kaplan 1949; Chandrasekhar 1964;
Merafina \& Ruffini 1989). Then, isolated ultrarelativistic
stars for which $P\sim\rho c^2$ are always dynamically unstable, while
stable existence of such configurations is possible, in principle, only
inside a box with fixed radius or fixed external pressure. Clearly, in
conditions of zero temperature there is no sense to discuss the
thermodynamical stability of the system. Therefore, the analysis made in
Section 3 and 4 by Chavanis (2002b) relates only to the dynamical
stability of such star (see also Yabushita 1974).

The loss of stability, dynamical as well as thermodynamical, may be found
from the linear series of equilibrium models, where the extremum of mass
in the appropriate curve $M(\rho _0)$ determines the appearance or
disappearance of unstable mode. For dynamical stellar stability this
``static" criterion was formulated by Zel'dovich (1963) and, for
thermodynamical instability, was used by Lynden-Bell \& Wood
(1968). Analysis of Chavanis (2002b) has shown that
dynamical instability of relativistic sphere in a box with fixed radius
happens exactly in the first maximum of the function $M(\rho_0)$.
Contrary to that, Yabushita (1974), who investigated dynamical
stability of gas spheres in a box with fixed external pressure in
general relativity (GR), had found the loss of stability in a point well
before the maximum of the curve $M(\rho_0)$. It is important to note that
not every kind of linear series of models may be used for investigation of
dynamical stability. In the case of an isolated star with a zero boundary
pressure it is necessary to have a fixed distribution of specific entropy
$S(m/M)$ along the series of models $M(\rho_0)$, where the first maximum
of $M(\rho_0)$ denotes the loss of stability relative to the global mode
without nodes (see Zel'dovich 1963). Analysis of dynamical
stability of stars in a box with fixed radius and fixed external pressure
had shown that only the first maximum of the curve $M(\rho _0)\vert_{r_e}$
corresponds to the loss of dynamical stability, but maximum on the curve
$M(\rho_0)\vert_{P_b}$ is situated after the point of the loss of
dynamical stability. The spheres in a box with constant external pressure
considered by Yabushita (1974) are not isolated from the
surrounding medium, which produces a work when the box is changing its
volume. So the comparison of two models with equal masses and different
central densities has no sense because these models have different
energies due to different volumes of the box. Therefore, the first maximum
on the sequence with constant $P_b$, considered by Yabushita does not
correspond to the onset of instability in spite of constant specific
entropy of the matter considered along the series. The coincidence of the
maximum of the curve $M(\rho_0)\vert_{P_b}$ with the point of the loss of
dynamical stability happens only at $q=0$, that formally corresponds to
zero external pressure and no external work.

In order to have a possibility to judge about the onset of dynamical
instability from the linear series of models in presence of constant
external pressure $P_b$ at the stellar boundary with a nondimensional
radius $\xi_b$ (usual Emden coordinates), we must take into account the
work of the external pressure and use the function

\begin{equation}
{\cal E}(\rho_0)=M+\frac{P_b V}{c^2},
\label{eq:21}
\end{equation}
\\
where $V$ is the total volume of the model, instead of $M(\rho_0)$. It is
easy to show, using Yabushita results and notations, that for these models
the extrema of the functions ${\cal E}(\rho_0)|_{P_b}$ or
$P_b(\xi_b)|_{\cal E}$ coincide exactly with the point of the loss of
stability for different modes, with increasing $\rho_0$ or $\xi_b$ (see
also Chavanis 2003).

\vspace{6pt}
\centerline{\it\noindent Star clusters with cutoff}

The second limit to which relativistic analysis of Chavanis (2002b) is
applied concerns ``isothermal" relativistic star clusters, with a local
constant temperature (Bisnovatyi-Kogan \& Zel'dovich 1969;
Bisnovatyi-Kogan \& Thorne 1970). Oppenheimer-Volkoff equations describing
the equilibrium of a relativistic cluster are given by

\begin{equation}
\left\{
\begin{array}{ll}
{\ds {dP\over dr} = - {G\over c^2}{(P+\rho c^2)(M_rc^2+4\pi Pr^3)\over
r(rc^2-2GM_r)}} & \\ & \\
{\ds {dM_r\over dr} = 4\pi r^2\rho ~,} &
\end{array}\right.
\label{eq:21a}
\end{equation}
\\
with $P(0)=P_0$ and $M_r(0)=0$. Pressure $P$ and total energy density
$\rho c^2$ are expressed as integrals in momentum space with the
distribution function (\ref{eq:25}), $M_r$ is the mass inside the lagrangian
radius $r$ (Bisnovatyi-Kogan et al. 1993, 1998). The Schwarzschild-type
metric was chosen

\begin{equation}
ds^2 = e^\nu c^2 dt^2 - e^\lambda dr^2 - r^2 \left({d\theta^2 +
sin^2 \theta d\varphi^2}\right)~,
\label{eq:21b}
\end{equation}
\\
where the metric coefficients are defined by the expressions

\begin{equation}
\left\{
\begin{array}{ll}
{\ds e^{\nu}=\exp\left({2\int_{r}^{\infty}{dP/dr\over P+\rho
c^2}~dr}\right)}
& \\ & \\
{\ds e^{\lambda}=\left({1-{2GM_r\over rc^2}}\right)^{-1}}. &
\end{array}\right.
\label{eq:21c}
\end{equation}
\\
In our calculations we have used the variable $W$ instead of $P$, which
leads to equilibrium equation (Merafina \& Ruffini 1989)

\begin{equation}
\ds{{dW\over dr} = - {G\over c^2}\left({1-\beta W\over\beta}\right)
{M_rc^2+4\pi Pr^3 \over r(rc^2-2GM_r)}} ~,
\label{eq:21d}
\end{equation}
\\
with the condition for integration $W(0)=W_0$. Here $\beta =T_R/mc^2$ and
$W$ is defined by (\ref{eq:23}). Therefore, each model is uniquely
determined by choosing the parameters $W_0$ and $T$ (or $\beta$).

These clusters are not in exact thermodynamical equilibrium, so the
investigation of the behavior of the linear series of models of such
clusters in a box does not give, rigorously speaking, correct results
about neither thermodynamical nor dynamical stability. Nevertheless at
$T\lesssim mc^2$, when the cluster is almost nonrelativistic and
gravitational potential $\varphi\ll c^2$, the local temperature is almost
constant all over the cluster and therefore the analysis of Chavanis
(2002b) can give a valid presentation about thermodynamical stability of
such system. In order to investigate dynamical instability of such
clusters, the equations for small perturbations should be solved or
variational principle may be used: this method, which is more complicated
in the relativistic case than in the Newtonian one, was derived, for
relativistic stars, by Chandrasekhar (1964) and Harrison et al. (1965);
for relativistic clusters, the variational principle derived by Ipser \&
Thorne (1968) and completed by Fackerell (1970) was used by
Bisnovatyi-Kogan \& Thorne (1970) for the investigation of dynamical
stability of the ``isothermal" clusters.

As it was noted by Lynden-Bell \& Wood (1968 and references
therein), the behavior of a star cluster in a box is related, in some
sense, to the isolated cluster with cutoff in the energy distribution.
Relativistic Maxwellian clusters with cutoff have been first introduced
by Zel'dovich \& Podurets (1965) and their dynamical stability
was studied by Bisnovatyi-Kogan et al. (1993, 1998).
Due to relativistic gravity, the loss of dynamical stability happens at
$\langle\gamma\rangle < \gamma_{cr}$ and the structure of the marginally
stable model strongly depends on the cutoff parameter. In the relativistic
cluster, the lower adiabatic index $\gamma < 5/3$ is connected with
relativistic motion of an ideal nondegenerate gas of stars: it is decreasing
with the increase of the temperature. Contrarily to that, a fully
degenerate highly non ideal nuclear matter is present in the interior of
a neutron star which may have even larger adiabatic index  $\gamma > 5/3$.

An analogy exists between the cutoff parameter $W_0$ used by Merafina \&
Ruffini (1989) and the quantity $v_1=m(\varphi_0-\varphi_e)/T$
used by Lynden-Bell \& Wood (1968), where $\varphi_0$ and $\varphi_e$
are the Newtonian gravitational potential in the center and at the edge of
the cluster respectively. In a full equilibrium cluster in the box with
the density distribution $\rho=\rho_0\exp\left[{m\varphi_0-m\varphi(r)}
\right]/T$, the quantity $-v_1$ represents the logarithm of the
ratio of densities in the center and at the edge of the cluster

\begin{equation}
-v_1={m(\varphi_e-\varphi_0)\over T}=\ln{\rho_0\over\rho_e}~.
\label{eq:22}
\end{equation}
\\
Note that density is a finite value at the outer boundary. The ratio
$\rho_0/\rho_e$ is also called ``density contrast". The value $W_0$ is
defined as

\begin{equation}
W_0=\left({\epsilon_{cut}\over T_r}\right)_{r=0}~,
\label{eq:23}
\end{equation}
\\
where $T_r =Te^{-\nu(r)/2}$ is the local thermodynamical temperature and
the constant $T$ is the temperature for an infinitely-remote observer. The
metric coefficient $\nu(r)$ has an usual meaning as in the Schwarzschild
metric (Landau \& Lifshitz 1962). It is easy to show (see
Bisnovatyi-Kogan et al. 1993) that, in Newtonian limit,
$W_0$ formally reduces to $-v_1$ in the first relation to the right side
of Eq.~(\ref{eq:22}) but, because of deviation from thermodynamical
equilibrium implied by the cutoff, the outer edge of the cluster
corresponds to zero density.

Thermodynamical stability of the isothermal cluster in Newtonian gravity
with a truncated distribution function may be characterized by the curve
$E_b(W_0)$ at constant $T$, where its maximum denotes the loss of
thermodynamical stability, $E_b=-Em/MT$ is a nondimensional specific
binding energy of the cluster, $E=T+U$ is the total energy of the cluster
(kinetic and gravitational energy). The value $Er_e/GM^2$, plotted in
Lynden-Bell \& Wood paper (1968) for an isolated equilibrium cluster in
a box as a functions of $-v_1$ at constant $r_e$, has the same meaning and
characterizes the thermodynamical stability of such cluster. The plot
$GM/r_eT$ as a functions of $-v_1$, also at fixed $M$ and $r_e$, going
along the sequence of models with varying $T$ determines, indeed, the
thermodynamical stability of the cluster in a thermal bath. The loss of
stability in such a cluster (at $-v_1=3.47$, with density contrast
$\rho_0/\rho_e =32.125$) happens before that of the isolated one (at
$-v_1=6.55$, with density contrast $\rho_0/\rho_e =708.61$). In the
isolated cluster perturbations are developed at constant energy but, for
the perturbed cluster in the bath, the temperature is preserved, so in the
last case the cluster occurs to be less stable.

The definition of the thermodynamical stability or instability of an open
cluster with a cutoff and zero density at the edge is, generically
speaking, senseless, because all such clusters are ``thermodynamically
unstable". The relaxation in these systems leads to approaching a local
Maxwellian distribution function without cutoff, equivalent to the
formation of a thermodynamically unstable isothermal gas sphere. Existence
of maxima on the curve $E_{b,T}(W_0)$ or, equivalently, on the curve
$E_{b,T}(\rho_0)$, where henceforth $E_b$ will simply indicate the
specific binding energy $E_b/N$, may signify, instead, the appearance of
an additional ``thermodynamically" unstable mode which is developed
without an increase of cutoff parameter and only due to a global
fluctuation of the parameter $T$. In reality, thermodynamical instability
of both types is governed by the so-called ``two-body relaxation
time" (Ambartsumyan 1938; Spitzer 1940)

\begin{equation}
\tau_b=8.8\cdot 10^5\sqrt{N{\bar R}^3\over {m/M_{\odot}}}{1\over
{\ln N-0.45}}~{\rm yr}~,
\label{eq:24}
\end{equation}
\\
where $m$ is the star mass, $\bar R$ is the radius of the cluster expressed
in parsec and $N$ is the total number of stars in the cluster, and therefore,
in this sense, all dynamically stable stellar clusters in space are
thermodynamically unstable. On the other hand, the local value of
$\tau_b$ is proportional to $1/n$, where $n$ is the number density of
stars, and the influence of cutoff is strong near the outer boundary,
where local $\tau_b$ has a maximal value. Moreover, after the point of the
loss of thermodynamical stability, the instability begins to develop also
in the central regions, where local $\tau_b$ has a minimal value, much
lower than at the edge of the cluster. Therefore, the loss of thermodynamical
stability determined by the curve $E_{b,T}(W_0)$ or $E_{b,T}(\rho_0)$ is
important for the cluster with a cutoff almost as well as for the cluster
in the box.

The analogy between our truncated clusters and isothermal cluster in the
box appears if we surround the truncated cluster by the box and let them
relax to thermal equilibrium. In this sense we exclude the global
thermodynamical instability of the open cluster and the curve
$E_{b,T}(\rho_0)$ or $E_{b,T}(W_0)$ gives informations about
thermodynamical stability of the cluster in a box with a radius equal to
the radius of the open truncated cluster. Comparison of Newtonian curve
of specific binding energy $E_b/N$ (see Fig.~1) for open clusters
considered in the paper of Bisnovatyi-Kogan et al. (1998)
with the corresponding one for clusters in a box, indicated in Fig.~2
of the paper of Lynden-Bell \& Wood (1968), shows a good
correspondence between first extrema of these curves which lay at
$-v_{1}=6.55$, for clusters in a box, and at $W_0 =6.42$, for open
clusters with truncated Maxwellian distribution function. This similarity
may be seen also from the comparison of these curves as a whole, where
approximate coincidence of two subsequent extrema is visible.

\begin{figure}[h]
\centerline{\includegraphics[scale=1.5]{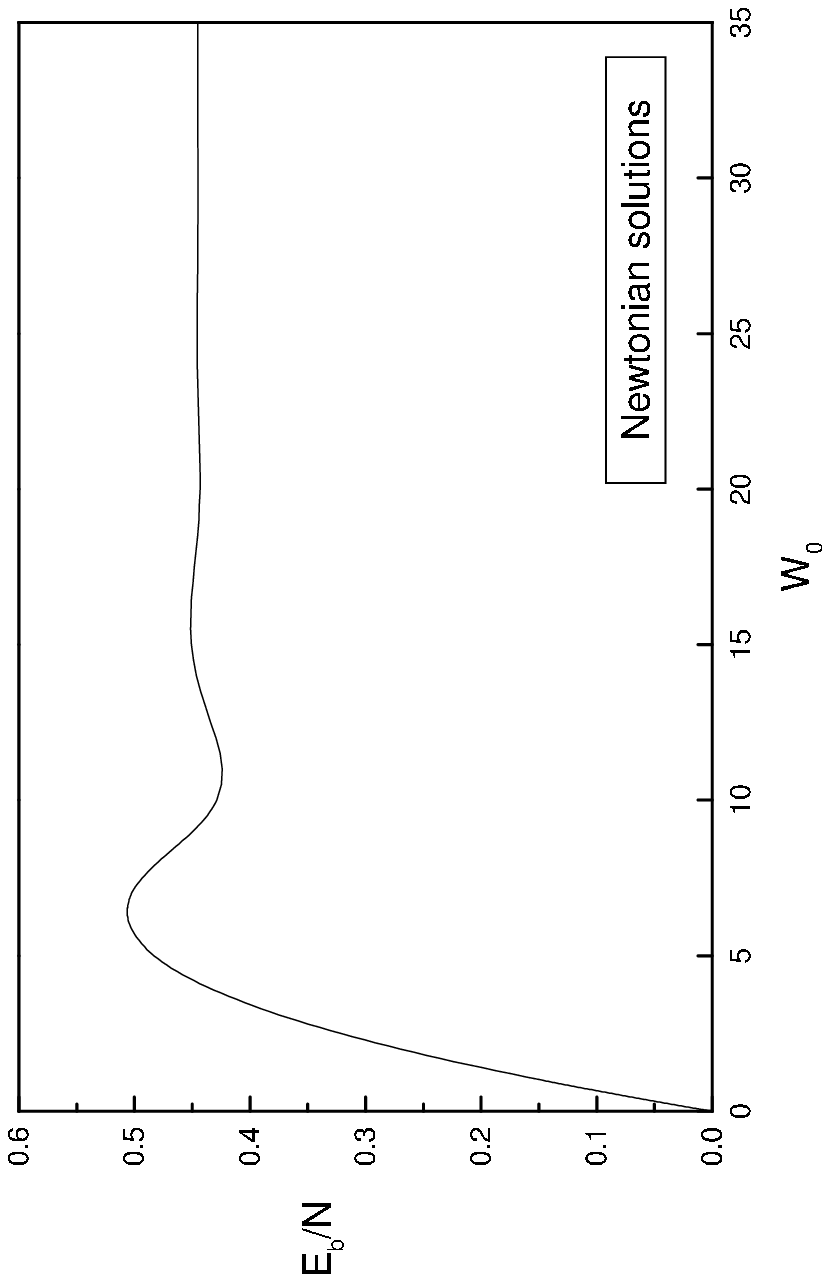}}
\figcaption[f1.eps]{Specific binding energy $E_b/N$ as a function of
$W_0$, representing the points of loss of thermodynamical stability of
different modes for Newtonian clusters with truncated Maxwellian
distribution. \label{fig1}}
\end{figure}

\placefigure{fig1}

\section{Relativistic and nonrelativistic oscillations in the curve
$E_b(\rho_0)$}
\centerline{\it\noindent Stability criteria for relativistic models}

The curve $E_b(\rho_0)$ characterizing the stability of the star shows a
strikingly similar oscillatory behavior for dynamical and thermodynamical
types of instability. This oscillatory behavior was first analyzed by
Dmitriyev \& Kholin (1963) in the curve $M(\rho_0)$ for cold
neutron stars. They have shown that the star loses its dynamical stability
relative to the global contraction in correspondence to the maximum of the
curve $M(\rho_0)$ and each new maximum and minimum leads to appearance of
a new unstable mode (see also Misner and Zapolsky 1964). The
detailed analysis of this curve was done by Harrison et al. (1965),
where the dependence $M(R)$ was also used: this choice is
even more useful for analysis of instability, because the behavior of the
spiral curve $M(R)$ permits to distinguish unambiguously between the
appearance of a new mode of instability or the removal of the unstable
mode, both of which may happen in the extremum of the curve $M(\rho_0)$.
The analysis of Chavanis (2002b), made for a simple relativistic
equation of state $P=q\epsilon$ in a box, is very similar to the
considerations of Dmitriyev \& Kholin (1963) and Harrison et al. (1965)
made for neutron stars.

The dependence of the total mass $M$ of the star on its total barion
number $N_b$, which has always a positive derivative and shows an angle on
the curve $M(N_b)$ corresponding to the extremum of the curve $M(\rho_0)$
obtained by Chavanis (2002b) for a star in the box, was also demonstrated
by Zel'dovich (1962b) for a cold neutron star.

Similar oscillations for the thermodynamical instability have been
obtained by Lynden-Bell \& Wood for an isothermal cluster in the box. In
the paper of Bisnovatyi-Kogan et al. (1998) the oscillatory behavior was
found in the curve $M_T(\rho_0)$ with a related spiral behavior of the
curve $M_T(\alpha)$, where the parameter $\alpha$ determines the cutoff of
the distribution function\footnote{Newtonian clusters with another type of
cutoff had been studied by Katz (1980).}

\begin{equation}
E\le mc^2 -\alpha T/2~,~~~~~f\sim e^{-E/T}~.
\label{eq:25}
\end{equation}
\\
This oscillatory (spiral) behavior, indicated by Bisnovatyi-Kogan {\it et
al.} (1998) for different fixed values of temperature $T$
(Newtonian case corresponds to $T\to 0$), shows appearance of modes of
``thermodynamical" instability of clusters with a cutoff, relative to
perturbations with heat redistribution inside the cluster and change
of the cutoff parameter $\alpha$. More detailed curves $M_T(\rho_0)$ and
$M_T(\alpha)$ are shown in Figs.~2 and 3, where results are represented in
nondimensional coordinates from Bisnovatyi-Kogan et al. (1998) with
using the same calculation scheme. Note that criterion for thermodynamical
stability based on the curve $M_T(\rho_0)$ works sufficiently well only
for large $T\gtrsim 0.1$. For smaller temperatures the criterion
based on the curve $E_{b,T}(\rho_0)$ should be used like in Newtonian
clusters (see next section).

\begin{figure}[h]
\centerline{\includegraphics[scale=1.5]{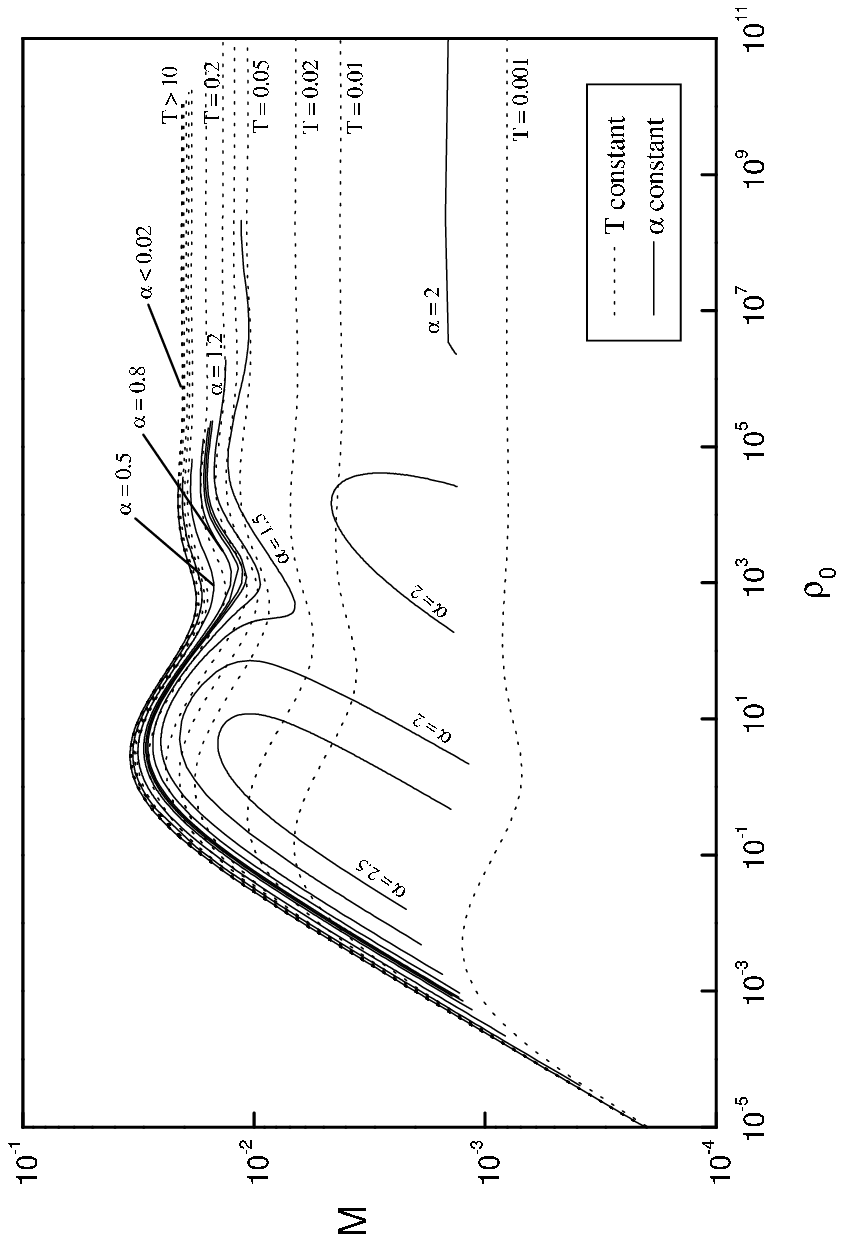}}
\figcaption[f2.eps]{Mass $M$ of equilibrium configurations in clusters
with a cutoff as a function of central density $\rho_0$ for different values
of temperature $T$ (dotted lines) and cutoff parameter $\alpha$
(continuous line). The curves representing dependence $M(\rho_0)$ at
constant $\alpha$ have several branches at $\alpha\gtrsim 1.5$; three
branches for $\alpha =2.0$ are represented. \label{fig2}}
\end{figure}

\placefigure{fig2}

Several approximate criteria had been suggested by Bisnovatyi-Kogan
et al. (1993) for investigation of dynamical stability of
relativistic clusters. All these criteria work almost equally well for
moderate values of $\alpha\lesssim 1.5$. At larger $\alpha$, the criterion
based on the evaluation of extremum of the curve $M_{\alpha}(\rho_0)$ is
not appropriate because of appearance of increasing number of loops (see
Fig.~2) which are connected with multiple intersections of the vertical
line $\alpha={\rm constant}$ with the spiral curves of Fig.~3, corresponding
to very low (zero) temperature. The first loop appears at $\alpha\simeq
1.5$, at $1.9\lesssim\alpha\lesssim 1.5$ there are two loops and,
in general, there is an even number of loops, except boundaries at
$\alpha\simeq 1.5~,~1.9~,~2.0~,~...~,~2.02$ at which there is an odd
number of loops $N_l=3~,~5~,~...~,~\infty$. At $\alpha<\alpha_{\infty}
\simeq 2.02$ there is a curve $M_{\alpha}(\rho_0)$ going to infinity and
at $\alpha>\alpha_{\infty}$ all curves $M_{\alpha}(\rho_0)$ are
represented only by loops at finite densities. Another important value is
$\alpha_{lim}\simeq 2.87$ so that there are no solutions for clusters
with a cutoff parameter $\alpha>\alpha_{lim}$ (see Bisnovatyi-Kogan et
al. 1998). The particular values of $\alpha$ given above
characterize the Newtonian curve $M_T(\alpha)$, at $T\to 0$. In fact, the
values $\alpha_{\infty}(T)$ and $\alpha_{lim}(T)$, changing for different
values of temperature $T$ (see center of spirals and maximum values of
$\alpha$ for each curve in Fig.~3), in the Newtonian limit of $T\to 0$
reach the particular values $\alpha_{\infty}=2.02$ and $\alpha_{lim}=2.87$
(see Bisnovatyi-Kogan et al. 1998).

\begin{figure}[h]
\centerline{\includegraphics[scale=1.5]{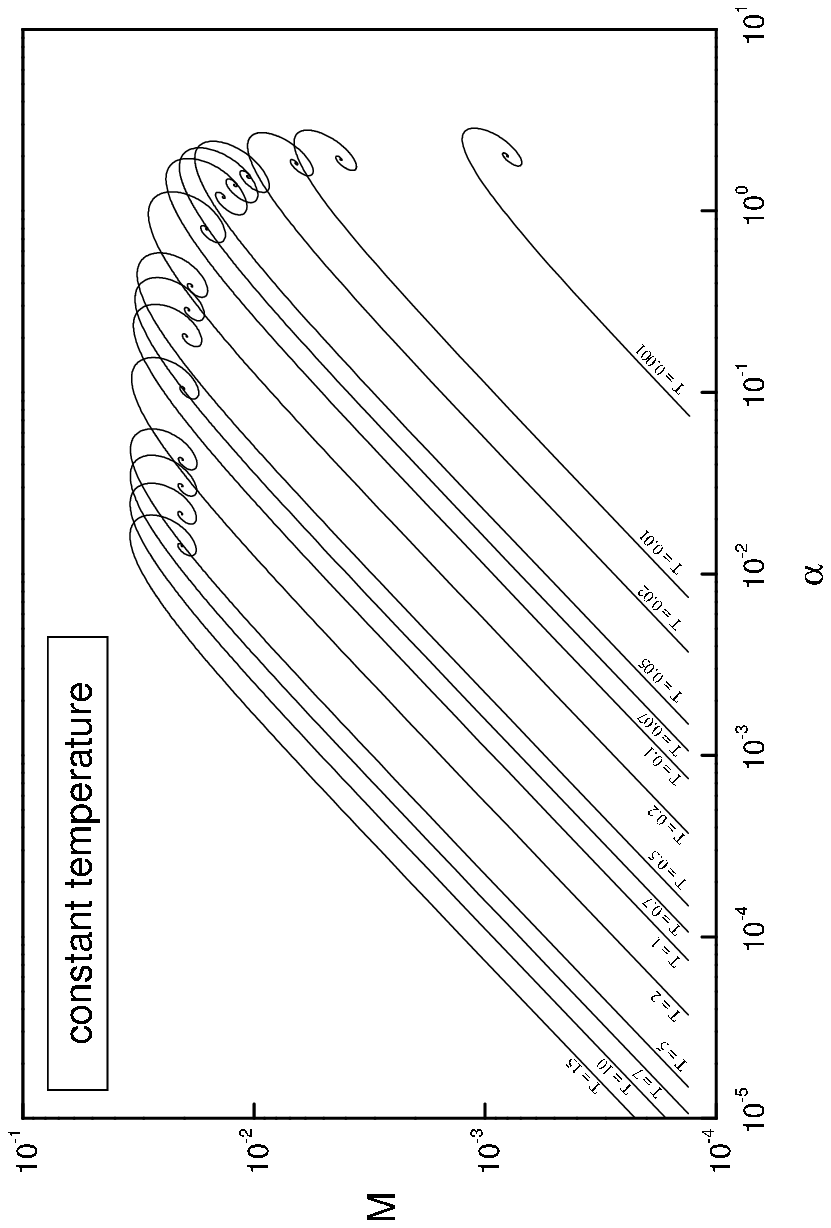}}
\figcaption[f3.eps]{Mass $M$ of equilibrium configurations in clusters
with a cutoff as a function of parameter $\alpha$ for different values of
temperature $T$. \label{fig3}}
\end{figure}

\placefigure{fig3}

The criteria based on evaluation of entropy and adiabatic invariants are
still valid, but their application is too complicated. Analysis made by
Bisnovatyi-Kogan et al. (1998) had shown that the most convenient
criterion of the dynamical stability of relativistic clusters should be
based on the investigation of the curves of dependence of the specific
binding energy of the cluster $E_b$ on the central density $\rho_0$, at
constant value of $W_0$. Note that binding energy of the relativistic
cluster is equal to the total energy of the cluster in Newtonian case,
where rest mass energy does not appear in the definition of $E$.

\begin{figure}[h]
\centerline{\includegraphics[scale=1.5]{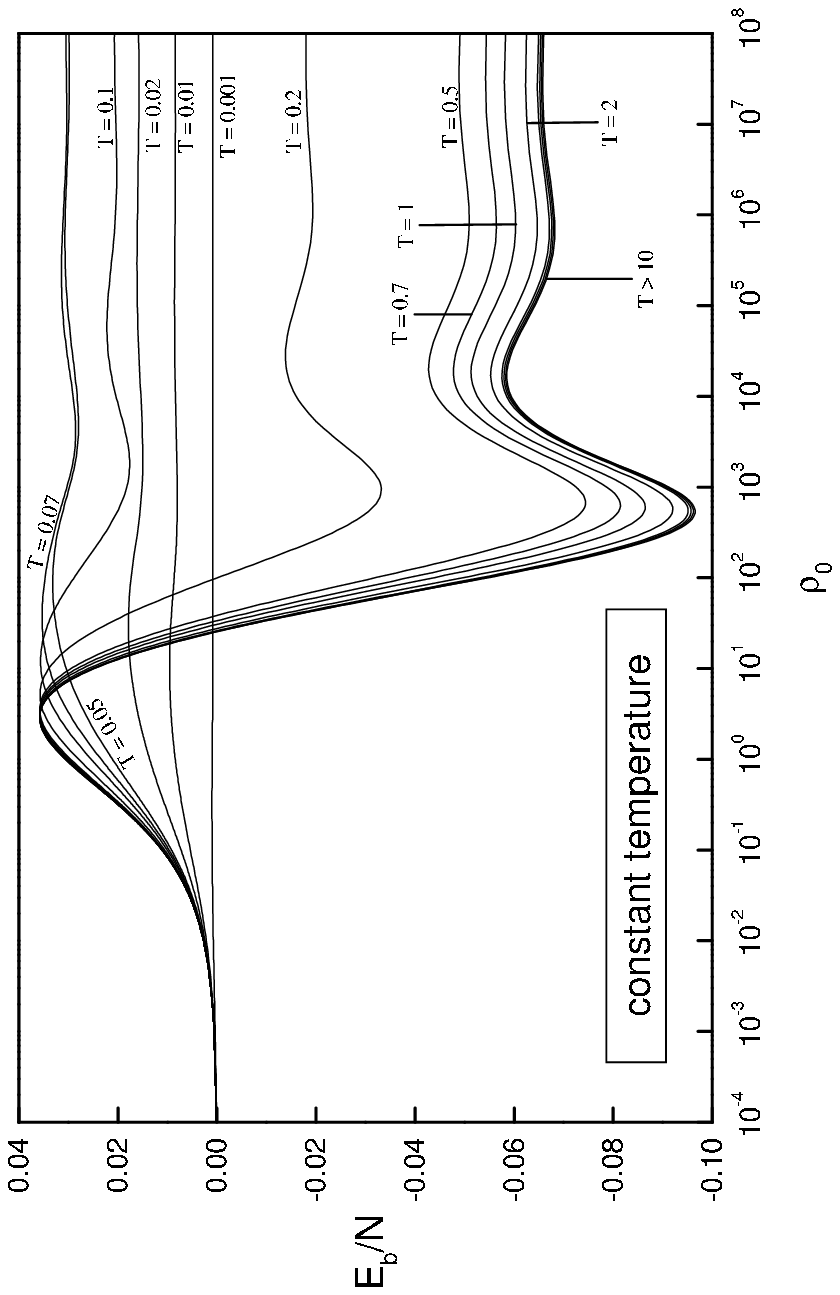}}
\figcaption[f4.eps]{Specific binding energy $E_b/N$ of equilibrium
configurations in clusters with a cutoff as a function of central density
$\rho_0$ for different values of temperature $T$. Each extremum corresponds
to appearance of new thermodynamically unstable modes. \label{fig4}}
\end{figure}

\placefigure{fig4}

\section{Numerical results}
\centerline{\it\noindent Turning point analysis for relativistic star
clusters}

To analyze the stability of relativistic clusters we need to calculate the
specific binding energy of the equilibrium models. We have calculated two
families of curves, which characterize dynamical and thermodynamical
stability of relativistic clusters with different cutoff parameters. The
curves $E_b(\rho_0)$ of specific binding energy at constant temperature
$T$ (Fig.~4) characterize the thermodynamical stability, while the curves
$E_b(\rho_0)$ at constant $W_0$ (Fig.~5) give information about dynamical
stability of the cluster. Relativistic expression of specific binding
energy is $E_b=(Nm-M)/Nm$, where $N$ is the total number of stars in the
cluster given by

\begin{equation}
{\ds N = 4\pi\int_{0}^{R}{nr^2 dr \over\sqrt{1-{2GM_r/rc^2}}}}
\label{eq:25a}
\end{equation}
\\
and $m$ is the mass of a single star (all stars have the same mass). The
number density $n$ is expressed as integral in momentum space with similar
calculation procedure used for obtaining the pressure $P$ and total energy
density $\rho c^2$ (Bisnovatyi-Kogan et al. 1993, 1998).

\begin{figure}[h]
\centerline{\includegraphics[scale=1.5]{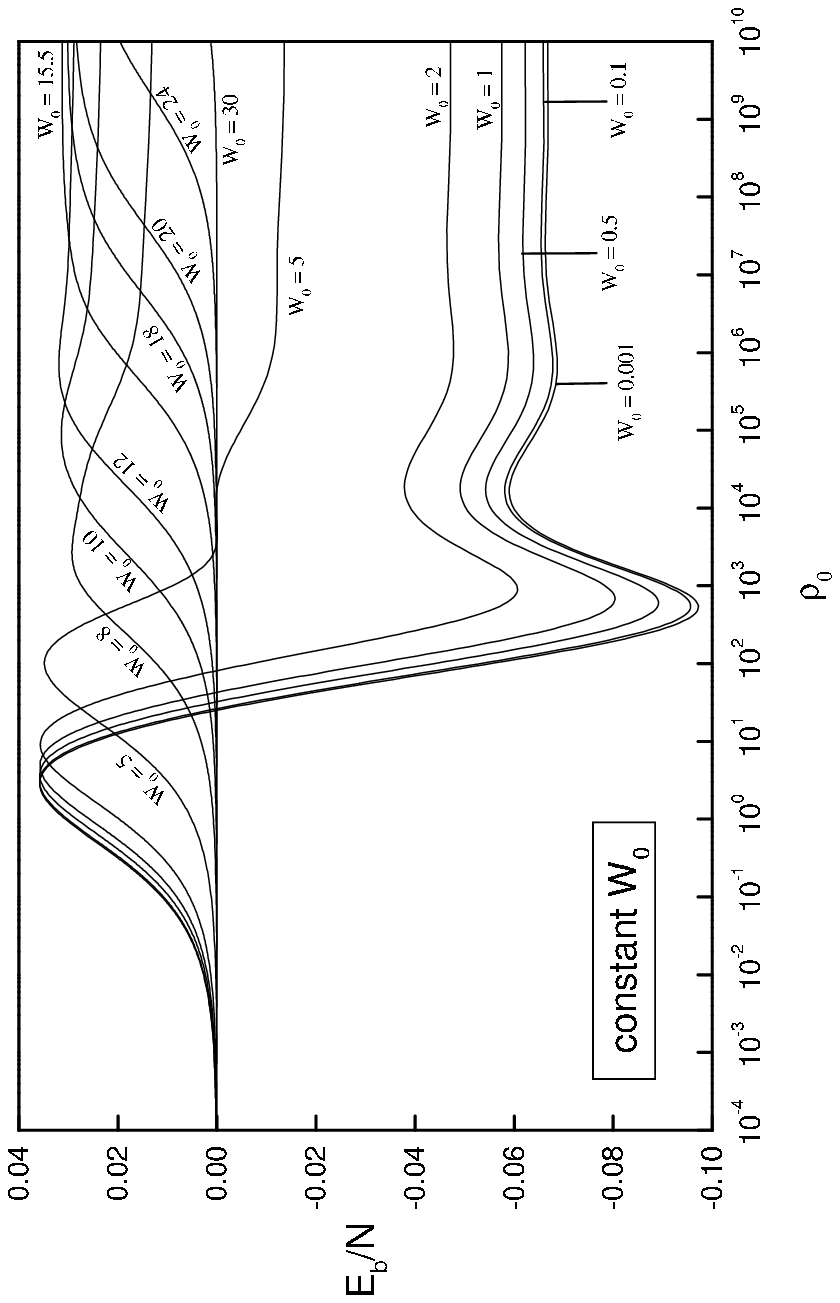}}
\figcaption[f5.eps]{Specific binding energy $E_b/N$ of equilibrium
configurations in clusters with a cutoff as a function of central density
$\rho_0$ for different values of parameter $W_0$. First maxima,
corresponding to loss of dynamical stability, are present only on curves
with $W_0\le 15.5$. Each extremum corresponds to appearance of new
dynamically unstable modes. \label{fig5}}
\end{figure}

\placefigure{fig5}

The temperature is increasing along each curve in Fig.~5, tending to a
finite constant value for large values of central density $\rho_0$. The
loss of stability, characterized by the first maximum, takes place only
for $W_0 \le 15.5$. In correspondence of this critical value $W_0\simeq
15.5$, the temperature, at large $\rho_0$, reaches a limiting value
$T_a=0.0635$. This means that no dynamic instabilities are present for
$T\lesssim 0.06$. At $W_0=16$, for example, the limiting temperature is
equal to $T_a=0.597$ and specific binding energy $E_b(\rho_0)$ increases
monotonously until the asymptotic value $E_{b,a}=0.0312$.

\begin{figure}[h]
\centerline{\includegraphics[scale=1.5]{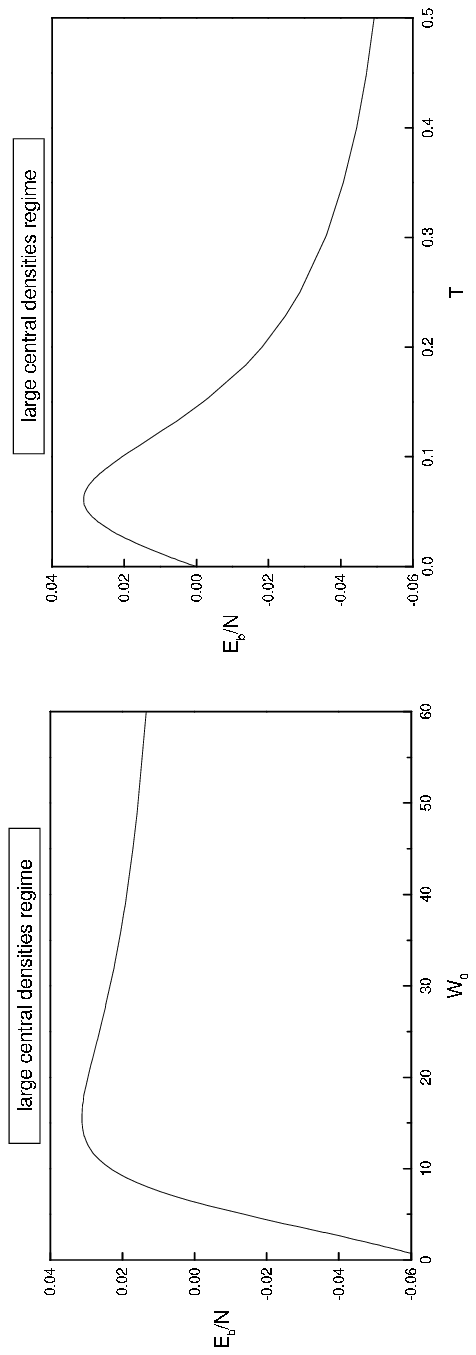}}
\figcaption[f6.eps]{Specific binding energy $E_b/N$ of equilibrium
configurations in clusters with a cutoff for very large central densities
$\rho_0$ and central redshifts $z_c$ as a function of $W_0$ (Fig.~6a, left
side) and $T_a$ (Fig.~6b, right side). The maximum, indicating the loss of
dynamical stability, corresponds to $W_0=15.8$ and $T\simeq 0.06$,
respectively. The limiting value of binding energy is $E_{b,a}=0.0312$.
\label{fig6}}
\end{figure}

\placefigure{fig6}

At large $\rho_0$, for models with very large central redshift $z_c$, there is
an asymptotic value $E_{b,a}$ of specific binding energy for each value of
$W_0$. Plotting the function $E_{b,a}(W_0)$ from Fig.~5 we obtain a more
precise boundary of the dynamical stability $W_{0,a}=15.8$. Due to
monotonic dependence of asymptotic (at large $\rho_0$) values of limiting
temperature $T_a$ on the parameter $W_0$, similar curve $E_{b,a}(T_a)$
from Fig.~4 shows the appearance of dynamically unstable clusters at $T_a
\gtrsim 0.06$. The limiting curves of specific binding energy $E_{b,a}(W_0)$
and $E_{b,a}(T_a)$ are represented in Figs.~6a and 6b respectively. In the
equilibrium configurations with very large central redshift, the temperature
is decreasing monotonously with the increase of $W_0$, as may be shown in
Fig.~7. Note that the curve $T_a(W_0)$ is approximated with a good precision
by the power-law relation

\begin{equation}
T_a=\frac{0.937}{(W_0)^{0.989}}~.
\label{eq:26}
\end{equation}

\begin{figure}[h]
\centerline{\includegraphics[scale=1.5]{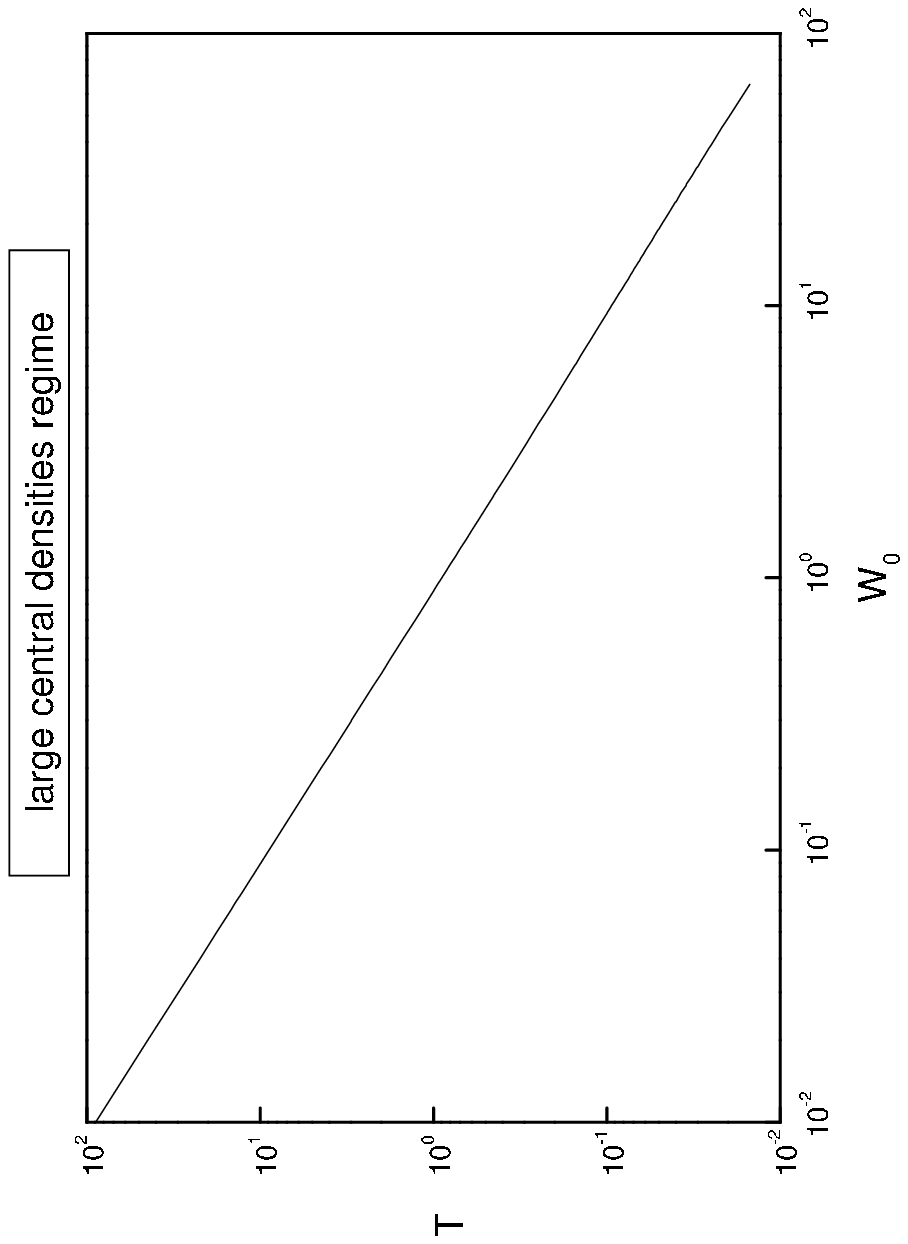}}
\figcaption[f7.eps]{Asymptotic values of temperature $T$ as a function of
$W_0$ in the limiting clusters with very high central redshifts. \label{fig7}}
\end{figure}

\placefigure{fig7}

Combining results of numerical investigation of dynamical and thermodynamical
stability are represented in Figs.~8,~9 and 10. In Figs.~8a and 8b dynamically
stable and unstable regions are represented in the planes $(T,\rho_0)$ and
$(T,z_c)$, respectively. The results plotted in the plane $(T,z_c)$ are
analogous to ones of the work of Merafina (1999). In Figs.~9a and
9b thermodynamically stable and unstable regions are represented in the
planes $(T,\rho_0)$ and $(T,z_c)$, respectively. Also in this case the
results plotted in the plane $(T,z_c)$ are analogous to ones of Merafina
(1999). Summary of numerical results on dynamical and thermodynamical
stability analysis is given in Figs.~10a and 10b, where different regions
are represented in the planes $(T,\rho_0)$ and $(T,z_c)$, respectively. It
is important to stress the coincidence of boundaries between dynamically
and thermodynamically stable and unstable configurations at large
temperatures. Using approximate criteria of dynamical stability we cannot
definitely judge if these boundaries coincide exactly or there is a small
difference between them, however the behavior of specific binding energy
$E_b/N$ let us enough confidence in this result.

\begin{figure}[h]
\centerline{\includegraphics[scale=1.5]{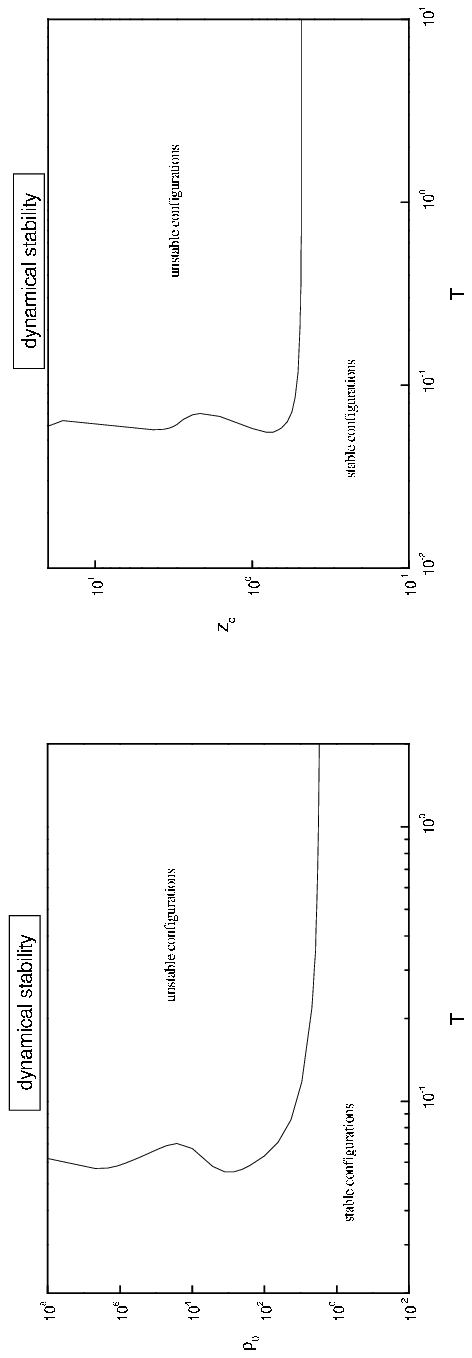}}
\figcaption[f8.eps]{Regions of dynamical stability and instability in the
plane $(T,\rho_0)$ (Fig.~8a, left side) and in the plane $(T,z_c)$
(Fig.~8b, right side). \label{fig8}}
\end{figure}

\placefigure{fig8}

Analysis of models with $\alpha$ approaching $\alpha_{\infty}$, when maximal
densities are expected, had shown that they have a very extended halo so,
even at very high densities, the cluster could remain to be dynamically stable
with local Newtonian properties. Similar situation should appear when
taking into account of degeneracy (Merafina \& Ruffini 1990). In
the Newtonian limit, only nonrelativistic degeneracy is expected, so the
situation is not changed qualitatively. The logarithmic curve $M(\rho_0)$
for a fully degenerate nonrelativistic gas is a monotonic line
$M\sim\rho_0^{1/2}$, like in a polytropic star with index $n=3/2$ and
$\gamma=5/3$, which does not show neither thermodynamical nor dynamical
instabilities. The thermodynamical instabilities on the curve $M_T(\rho_0)$
appear only at finite temperatures.

\begin{figure}[h]
\centerline{\includegraphics[scale=1.5]{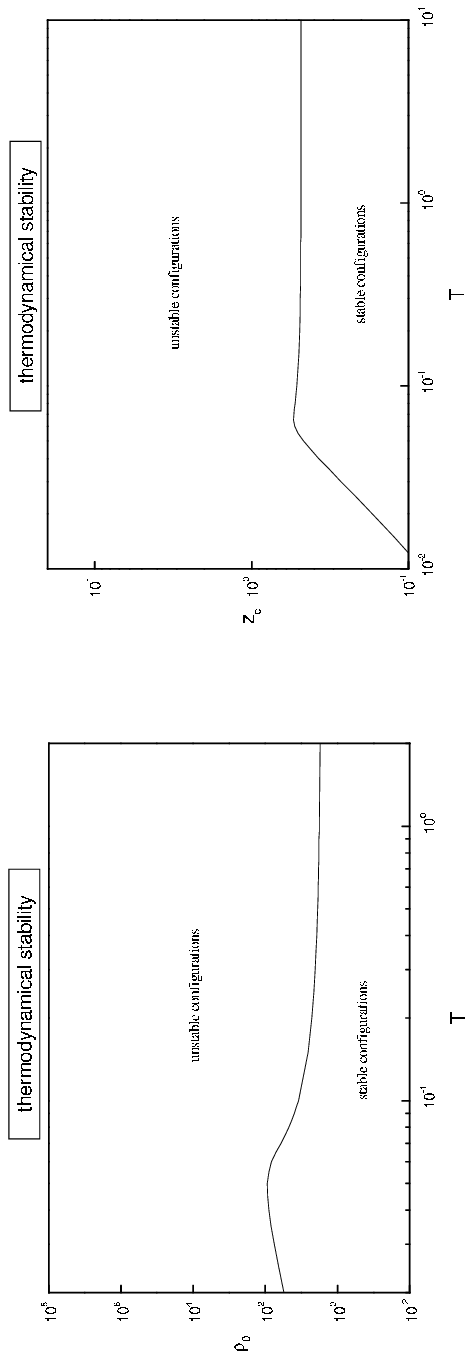}}
\figcaption[f9.eps]{Regions of thermodynamical stability and instability
in the plane $(T,\rho_0)$ (Fig.~9a, left side) and in the plane $(T,z_c)$
(Fig.~9b, right side). \label{fig9}}
\end{figure}

\placefigure{fig9}

Consideration of fully degenerate ultrarelativistic particles with equation
of state $P=q\epsilon$, made by Chavanis (2002b), has sense only
for a limited box because these configurations are dynamically unstable
with an open outer boundary. The curve $E_b(\rho_0)$ for a given size of
the box gives results about the onset of dynamical instability in such
configurations and has an analogy with the corresponding curve at constant
$W_0$.

\begin{figure}[h]
\centerline{\includegraphics[scale=1.5]{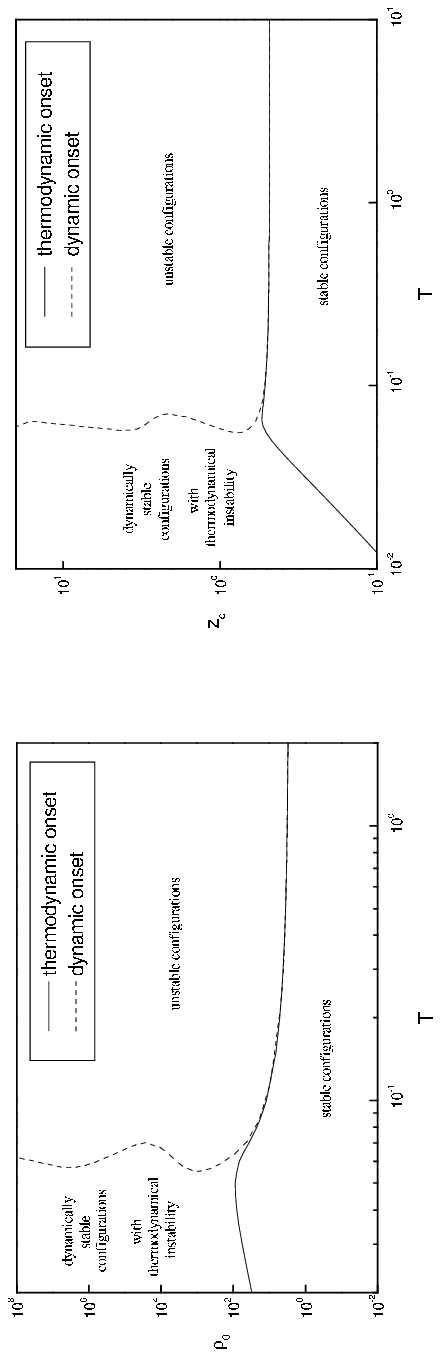}}
\figcaption[f10.eps]{Regions of dynamical and thermodynamical stability
and instability in the plane $(T,\rho_0)$ (Fig.~10a, left side) and in the
plane $(T,z_c)$ analogous to results of Merafina in 1999 (Fig.~10b, right
side). \label{fig10}}
\end{figure}

\placefigure{fig10}

In the problem of stellar stability the curve $M_S(\rho_0)$, where $S$ is
the specific entropy of an isentropic star, is used in the static criterion
and its maximum determines the boundary of dynamical stability of a
star (Zel'dovich 1963). The curve $M(\rho_0)$ at constant
``entropy" was also used for an approximate estimation of the boundary of
dynamical stability of truncated stellar cluster (Bisnovatyi-Kogan et
al. 1993), but such approach is very cumbersome. Using the same
curve at constant $\alpha$ or $W_0$ is much easier and give very close
results. It happens however, that at large values of $W_0\gtrsim 11$ or at
large enough $\alpha\gtrsim 2$, the curve $M(\rho_0)$ is not valid
anymore for the estimation of the dynamical stability. The corresponding
curve at constant $\alpha$ becomes discontinuous and irrelevant. Using the
curve $E_b(\rho_0)$ instead of $M(\rho_0)$ permits to make such estimation
in all range of values of the parameter $W_0$. We have obtained numerically
that at $W_0\lesssim 11$ maxima of the curves $E_b(\rho_0)$ and
$M(\rho_0)$ coincide exactly but, at larger $W_0$, maximum of the curve
$M(\rho_0)$ disappears while the maximum of the curve $E_b(\rho_0)$ is
remaining and shifting to infinite central density at $W_0=15.8$. So
that all models at larger $W_0$ are dynamically stable. Similar relation
between $E_b(\rho_0)$ and $M(\rho_0)$ takes place for the curves at
constant $T$. At large values of $T\gtrsim 1$ maxima of these curves
coincide and the curves themselves tend to an asymptotic behavior
(see Figs.~2 and 4). At smaller temperatures the maximum of $E_b(\rho_0)$
is shifting to larger densities (see Fig.~4) and maximum of $M(\rho_0)$
moves in the opposite side of smaller densities. Following Lynden-Bell \&
Wood (1968) we accept that loss of thermodynamical stability is
connected with the maximum of the curve $E_{b,T}(\rho_0)$.

\section{Conclusions}
The visible similarity between the oscillations of the curve $E_b(\rho_0)$
for relativistic clusters and the Newtonian isothermal stellar cluster in
the box may have a different physical nature: in the first case there are
two types of oscillations, reflecting the loss of dynamical and
thermodynamical stability, while the cluster in the box is dynamically
stable and oscillations are connected only with the onset of thermodynamical
instability leading to gravothermal catastrophe.

The analogy between the open cluster with a cutoff in the Maxwellian energy
distribution function and an isothermal cluster in a box is not complete
because the first one is thermodynamically unstable everywhere while the
second one is only after the Antonov's point, at density contrast
$\rho_0/\rho_e\gtrsim 709$. Nevertheless, both kinds of oscillations are
present in the structure of relativistic clusters with a cutoff when we
consider clusters with different cutoff parameter $\alpha$ and a variable
value $W_0$, which plays the role of the ratio $\rho_0/\rho_e$ (density
contrast) or $-v_1$ for the cluster in a box. The Figs.~4 and 5 illustrate
the behavior of both oscillations for the function $E_b(\rho_0)$, where
oscillations of the curves at constant $T$ add thermodynamically unstable
modes and oscillations of the curves at constant $W_0$ give a picture of
the dynamical stability of the cluster. Calculations show that Newtonian
truncated clusters lose their thermodynamical stability at $W_0=6.42$,
very close to the value $-v_1=6.55$ for clusters in a box.

We have obtained that only the curve $E_b(\rho_0)$ at constant $T$ is
valid for the estimation of thermodynamical stability, while the
corresponding curve $M(\rho_0)$ may be misleading. In the case of dynamical
stability both curves give identical results at lower values of
$W_0\lesssim 11$ while, at larger ones, only the curve $E_b(\rho_0)$ may be
used.

\vfill\eject

\clearpage

\clearpage

\clearpage

\clearpage

\clearpage

\clearpage

\clearpage

\clearpage

\clearpage

\end{document}